\documentclass[iop]{emulateapj-rtx4}
\usepackage{amsmath}
\usepackage[usenames,dvipsnames]{xcolor}

\usepackage[normalem]{ulem}
\usepackage{cancel}

\newcommand{\MJup}{M_{\rm Jup}}
\newcommand{\Msolar}{{\rm M_{\odot}}}   
\newcommand{\Rt}{R_{\rm t}}
\newcommand{\cs}{c_{\rm s}}
\newcommand{\vb}{v_{\rm b}}
\newcommand{\vf}{v_{\rm f}}
\newcommand{\Mstar}{M_{\star}}
\newcommand{\Mdisc}{M_{\rm disk}}
\newcommand{\tcoll}{t_{{\rm coll},ij}}
\newcommand{\torb}{t_{\rm orb}}
\newcommand{\icarus}{Icarus}
\newcommand{\Sci}{Science}

\shorttitle{Grain growth in Brown Dwarf disks}
\shortauthors{Meru et al.}

\begin{document}


\title{Growth of grains in Brown Dwarf disks}


\author{Farzana Meru\altaffilmark{1}}
\affil{Institut f\"ur Astronomie, ETH Z\"urich, Wolfgang-Pauli-Strasse 27, 8093 Z\"urich, Switzerland}
\email{farzana.meru@phys.ethz.ch}


\author{Marina Galvagni}
\affil{Institute of Theoretical Physics, Universit\"at Z\"urich, Winterthurerstrasse 190, 8057 Z\"urich, Switzerland}

\and

\author{Christoph Olczak}
\affil{Astronomisches Rechen-Institut (ARI), Zentrum f{\"u}r Astronomie Universit{\"a}t Heidelberg, M{\"o}nchhofstrasse 12-14, 69120 Heidelberg, Germany}
\affil{Max-Planck-Institut f{\"u}r Astronomie (MPIA), K{\"o}nigstuhl 17, 69117 Heidelberg, Germany,}
\affil{National Astronomical Observatories of China, Chinese Academy of Sciences (NAOC/CAS), 20A Datun Lu, Chaoyang District, Beijing 100012, China}




\begin{abstract}
We perform coagulation and fragmentation simulations using the new physically-motivated model by \citealp{Garaud_vel_pdf} to determine growth locally in brown dwarf disks.  We show that large grains can grow and that if brown dwarf disks are scaled down versions of T Tauri disks (in terms of stellar mass, disk mass and disk radius) growth at an equivalent location with respect to the disk truncation radius can occur to the same size in both disks.  We show that similar growth occurs because the collisional timescales in the two disks are comparable.  Our model may therefore potentially explain the recent observations of grain growth to millimetre sizes in brown dwarf disks, as seen in T Tauri disks.

\end{abstract}





\section{Introduction}

Recent observations of brown dwarf disks suggest that grains may grow in such disks, just as in T Tauri disks.  In addition, there are suggestions that the former may be scaled-down versions of the latter: \citealp{Apai_BDdiscs_2micron} presented mid-infrared observations of the disk around CFHT-BD-Tau 4 and concluded that intermediate-sized ($\approx 2 \mu m$) dust grains dominated the disk atmosphere, suggesting that the initial phase of grain growth had occurred.  Grain growth, and the resulting midplane settling that follows, naturally explains why brown dwarf disks do not appear to be as flared as expected from disk models in vertical hydrostatic equilibrium with gas and dust being well-mixed \citep[e.g.][]{Apai_BDdiscs_2micron,Walker_BDdisc_structure}.

\citealp{Apai2005_BDdiscs_Sci} obtained mid-infrared spectra of six brown dwarf disks and showed broad emission features indicating the presence of larger grains ($\approx O(1) \mu$m).  They also suggested that dust processing is independent of stellar properties and suggested that the first steps of planet formation in brown dwarf disks is similar or identical to that around low- and intermediate-mass stars.

\citealp{Klein_BDdiscs_mm} performed a millimetre survey of disks around brown dwarfs and obtained a range of masses for the disks around CFHT-BD-Tau 4 and IC 348 613.  Though there were some uncertainties in the disk and brown dwarf masses, they suggested that the disk to primary mass ratios were consistent with the equivalent in a T Tauri star-disk system.

\citealp{Scholz_BDdiscs_scaleddown_TT} performed a millimetre survey of brown dwarf disks in Taurus.  Combined with mid-infrared data they suggested that brown dwarfs seem to have disks which are scaled down versions of T Tauri disks.  Furthermore, \citealp{Mohanty_BD_VLMS_discs} performed SCUBA-2 850 $\rm \mu m$ observations of disks around very low mass stars and brown dwarfs in Taurus and the TW Hydrae Association and showed that the apparent disk to star mass ratio was roughly constant.

Recently, observations of large-sized objects (mm-sized or larger) in brown dwarf disks \citep{Bouy_BDdiscs_mm,Ricci_mm_cm_BD_rhoOph,Ricci_mm_cm_2MASS,Mohanty_BD_VLMS_discs} are questioning how growth in such disks occurs.  However, the theory of grain growth in brown dwarf disks and how similar it is to that in T Tauri disks has lagged behind observations.  \citealp{Pinilla_BD_discs} performed coagulation and fragmentation simulations to determine under what conditions growth may occur in brown dwarf disks.  They found that a particular combination of disk and collision parameters (surface mass density profile, disk outer radius, fragmentation velocity and turbulence parameter) was needed to match the observables (flux and spectral index at millimetre wavelengths) in the case where the radial drift velocity and radial motion was not considered, as well as when radial drift was included (though in this case pressure inhomogeneities of a particular strength were also needed).

Here we take a more general approach towards understanding growth in such disks focussing on how this compares to growth in T Tauri disks, with the latter being scaled-up versions of the former.  We use a new physically-motivated approach developed by \citealp{Garaud_vel_pdf}.  While this model uses a probability distribution function (PDF) for the velocity of the dust aggregates \citep[a technique also used by][]{Okuzumi2011,Galvagni2011,Windmark_vel_pdf}, this approach \emph{crucially} separates the velocity contributions into deterministic (those that are directional, i.e. radial drift, azimuthal drift and vertical settling) and stochastic (those that do not have a set direction, i.e. turbulence and Brownian motion) components.  

We show that growth may potentially occur in brown dwarf disks simply by considering that a particle may have a distribution of velocities and the fact that collisions between low  and high mass aggregates can withstand destruction better than collisions between equal-mass aggregates.  We also show that growth in brown dwarf disks can in principle occur to the same sizes as in T Tauri disks, consistent with recent observations.  We describe our method in Section~\ref{sec:method}.  We then present our simulations and results in Section~\ref{sec:sim_results}.  Finally, we discuss our results and present our conclusions in Sections~\ref{sec:disc} and~\ref{sec:conc}, respectively.

\section{Method}
\label{sec:method}

We use the model developed by \citealp{Garaud_vel_pdf} which simulates the local coagulation and fragmentation of dust aggregates to determine the evolution of the particle size distribution function using the Smoluchowski equation \citep{Smoluchowski1916,Melzak1957}.  The key feature compared to previous models is that not only is a particle's velocity modelled using a PDF, but crucially, the deterministic (i.e. directional) and stochastic (i.e. non-directional) velocities are separated.

This method uses a one-dimensional Gaussian distribution of velocities for each particle in each direction with a mean given by the directional velocities (radial drift, azimuthal drift or vertical settling) and the standard deviation determined by the stochastic components (turbulence and Brownian motion).  The one-dimensional PDF of the \emph{relative velocity} between two particles in any one direction is determined (equation 24 of \citealp{Garaud_vel_pdf}) and used to produce a three-dimensional velocity PDF (equation 26 of \citealp{Garaud_vel_pdf}).

We assume that collisions with velocities lower than the bouncing velocity, $\vb$, result in sticking while those with velocities higher than the fragmentation velocity, $\vf$, break apart if their mass ratio is smaller than a particular value (called the \emph{Mass Transfer} parameter) and coagulate if it is larger.  Physically this means that collisions between unequal-sized aggregates are more likely to lead to growth while equal-sized aggregates are more likely to fragment - a result shown by laboratory experiments \citep{Wurm_25m/s_impacts,Teiser_Wurm_highVcoll} and simulations (Meru et al, submitted).  To easily compare with previous work \citep{Windmark_vel_pdf,Garaud_vel_pdf} we choose $\vb = 5 \rm cm/s$ and $\vf = 100 \rm cm/s$.  For collisions with velocities $\vb < v < \vf$, the aggregates bounce.  The particle volume density is $1 \rm g/cm^3$

The minimum and maximum sizes in our simulations are $1 \times 10^{-5}$ and $1 \times 10^{5} $cm, respectively, and we use 120 logarithmically-spaced size bins (i.e. 12 bins per decade in size or equivalently, 4 bins per decade in mass).  The initial particle distribution is a Gaussian centred around a size of $1 \times 10^{-4}$cm with a width of $1 \times 10^{-5}$cm. 

The simulations are evolved until the largest particle size reaches a steady-state.  The orbital time for the T Tauri disk is $\approx 3.5$ times longer than for the brown dwarf disk (at equivalent locations with respect to the truncation radius).  Therefore, these simulations have been run for 3.5 times longer.

We note that a number of numerical aspects can affect the number density of particles at a given size, the maximum particle size or the evolution of the number density distribution, including the resolution \citep{Garaud_vel_pdf}, the minimum particle size and the initial distribution.  Therefore, in this Letter we focus on the \emph{qualitative} comparison of the \emph{final} surface density distribution in brown dwarf and T Tauri disks.

\section{Simulations \& Results}
\label{sec:sim_results}

We perform local simulations with brown dwarf disk parameters guided by the observational data from \citealp{Ricci_mm_cm_BD_rhoOph}.  We determine the growth of aggregates at radii $R  = 2$, 5 and 10au from the star with mass $\Mstar = 60 \MJup$.  The disk surface mass density has the form

\begin{equation}
\Sigma = \frac{\Mdisc}{2 \pi R \Rt} e^{-\frac{R}{\Rt}},
\label{eq:Sigma}
\end{equation}
where $\Mdisc = 4 \times 10^{-4} \Msolar$ is the disk mass, $\Rt = 15$au is the truncation radius (the radius beyond which the surface mass density rapidly drops off), and the sound speed is given by

\begin{equation}
\cs = c_{\rm s,au} \left ( \frac{R}{1 \rm au} \right )^{-\frac{1}{4}},
\label{eq:cs}
\end{equation}
where $c_{\rm s,au} = 6.5 \times 10^4$cm/s is the sound speed at 1au.  This is equivalent to temperatures of 83, 52 and 37 K at $R = 2$, 5 and 10au, respectively.  We assume a \citealp{SS_viscosity} type $\alpha$-viscosity with a turbulence parameter $\alpha = 1 \times 10^{-4}$ and Reynolds number $\rm Re = 10^8$ (as in \citealp{Garaud_vel_pdf} and \citealp{Windmark_vel_pdf}).  We use a mass transfer parameter, $\phi = 500$ (note that this is 10 times larger and hence a stricter condition for growth than used previously by \citealp{Garaud_vel_pdf} and \citealp{Windmark_vel_pdf}).

The T Tauri disk is set up to surround a $1\Msolar$ star and has the same disk to star mass ratio as the brown dwarf disk.  We set the truncation radius ($\Rt = 90$au) according to $\Mdisc \propto \Rt^{1.6}$, a relation determined observationally by \citealp{Andrews_Md_Rt} for T Tauri disks in Ophiuchus.  Therefore our simulated T Tauri disk is a ``scaled-up'' version of the brown dwarf disk in terms of the stellar mass, disk mass and disk radius.  The disk has the same surface mass density and temperature profiles as the brown dwarf disk and the same absolute temperature at 1au (though we perform a test to illustrate the limited effect of temperature on the growth).  The simulations are performed at $R = 12$, 30 and 60au, i.e. 2/15, 1/3 and 2/3 of the distance to the truncation radius, as simulated in the brown dwarf disk.

\begin{figure*}
\centering
\includegraphics[width=0.66\columnwidth]{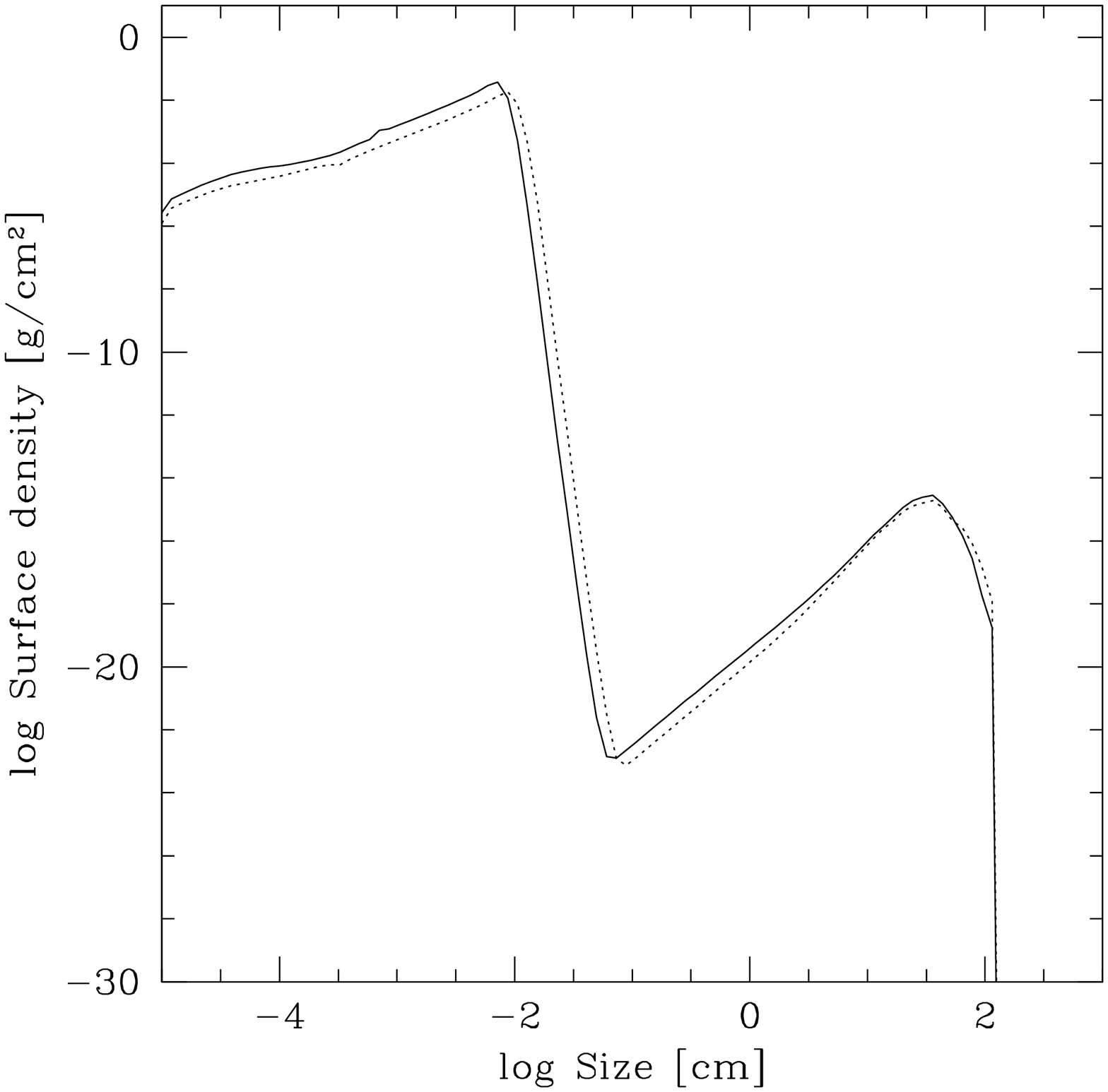}
\includegraphics[width=0.66\columnwidth]{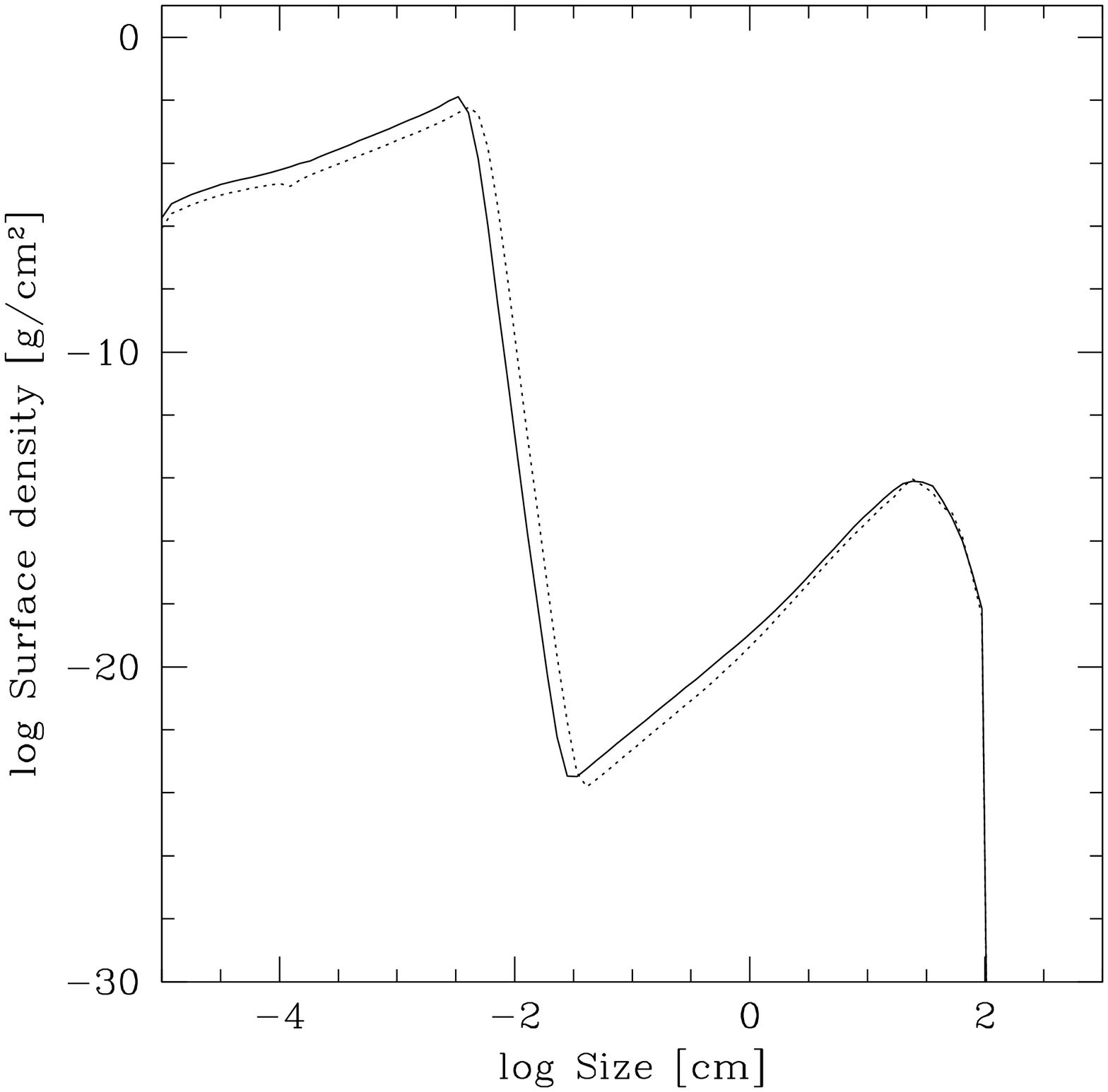}
\includegraphics[width=0.66\columnwidth]{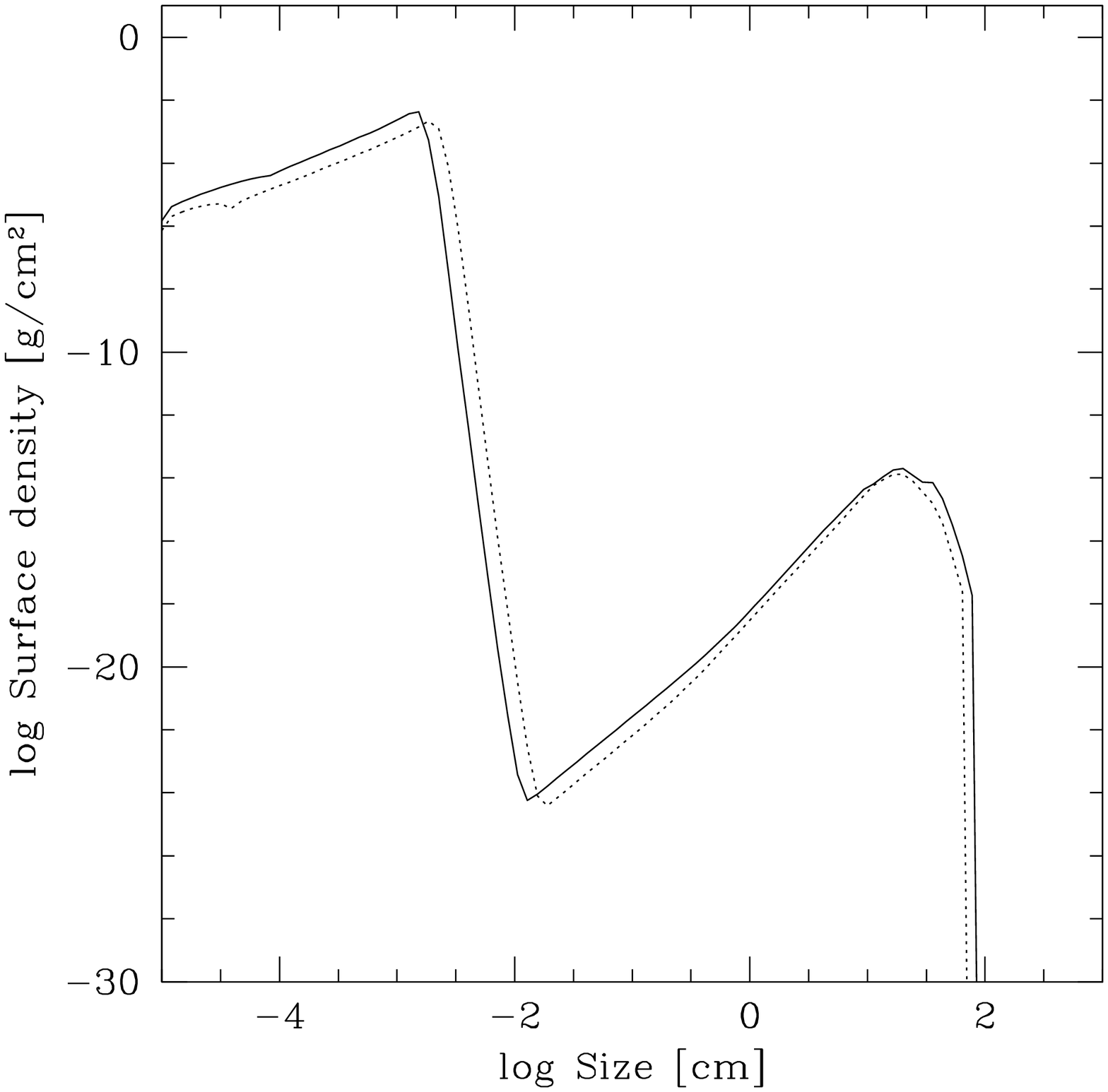}

\vspace{0.3cm}

\includegraphics[width=0.67\columnwidth]{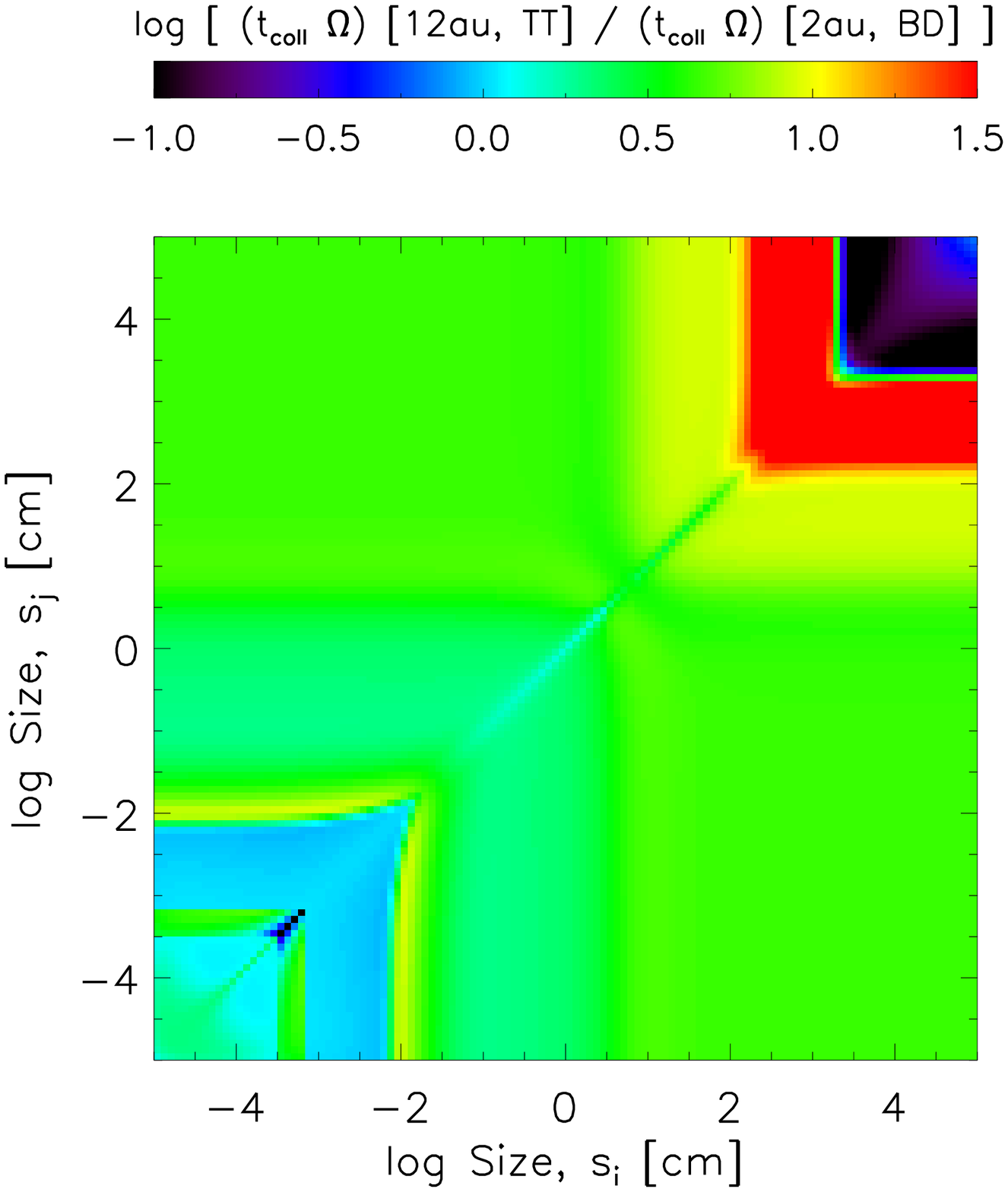}
  \includegraphics[width=0.67\columnwidth]{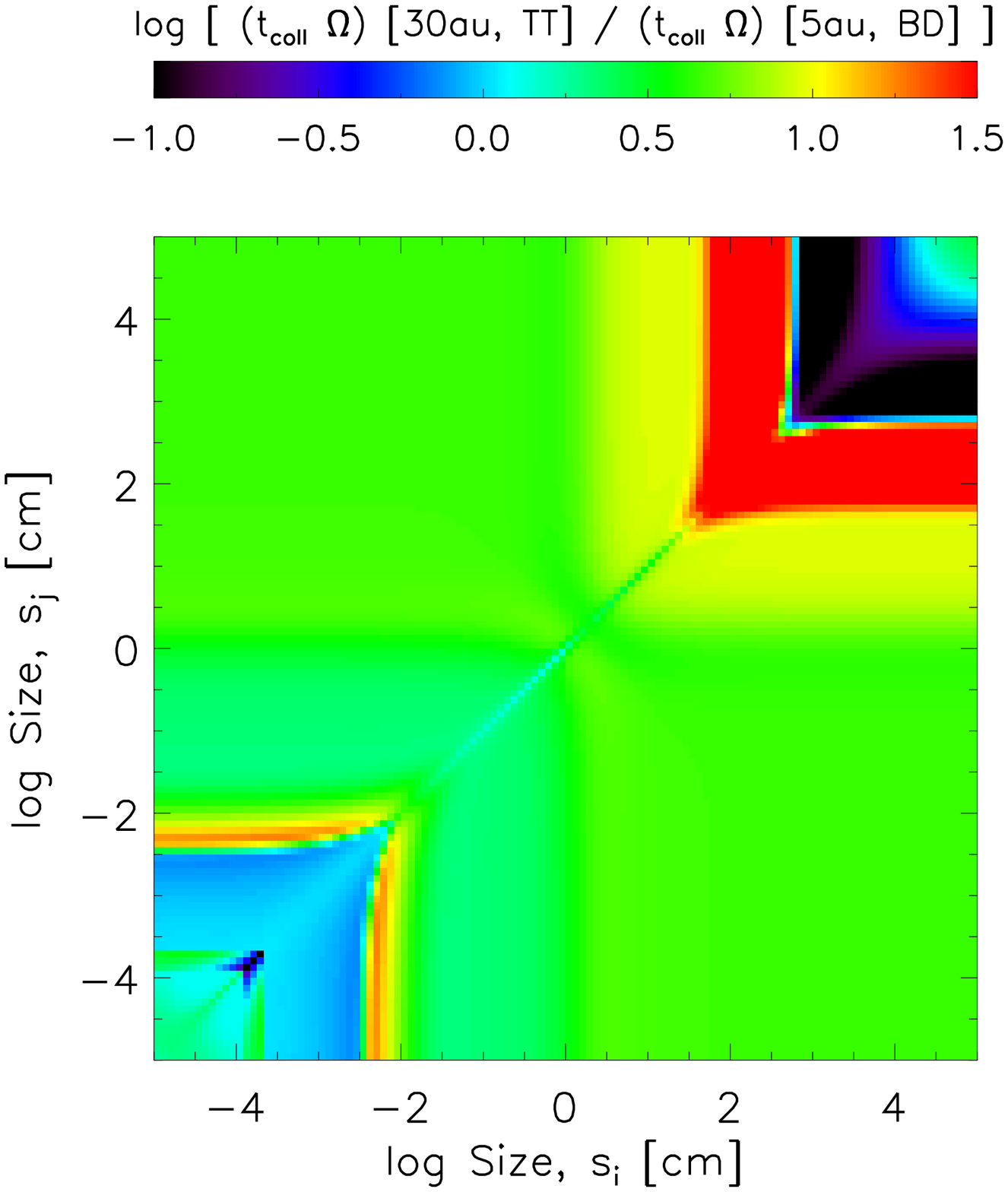}
  \includegraphics[width=0.67\columnwidth]{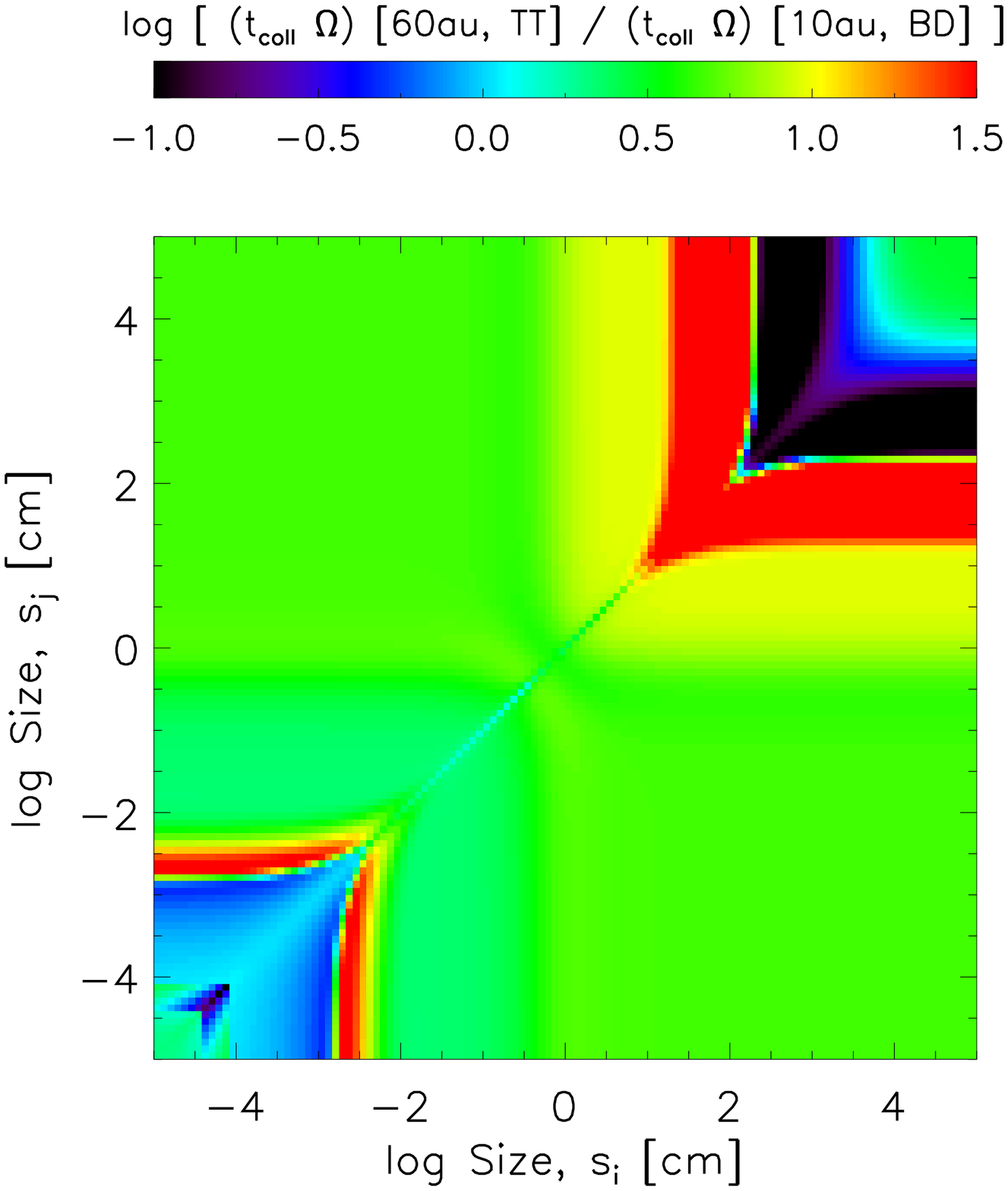}
 \caption{Surface mass density (top panel) of aggregates against particle size for the brown dwarf (solid line) and T Tauri disks (dotted line) at radii of 2/15 (left panel), 1/3 (middle panel) and 2/3 (right panel) of the distance to the truncation radius.  In the brown dwarf disk these are equivalent to 2au, 5au and 10au, respectively, and 12au, 30au and 60au, respectively, in the T Tauri disk.  The lower panel shows the ratio of the collisional timescales (in units of the orbital timescales) of the simulations in the panels above.  Growth at equivalent locations in the two disks occurs to the same size because the collisional timescales are similar.}
\label{fig:sigma_tcoll_BD_TT}
\end{figure*}

The existence of \emph{Mass Transfer} causes two particle populations to emerge (\citealp{Windmark_lucky_ptcl,Garaud_vel_pdf}).  These characteristic graphs consist of a small size population and a second population of larger-sized aggregates, resulting in two peaks in the distribution (Figure~\ref{fig:sigma_tcoll_BD_TT}, top panel).  We define the maximum size to be the largest size before the surface mass density of aggregates drops off completely, i.e. the second peak in the distribution.

Figure~\ref{fig:sigma_tcoll_BD_TT} (top panel) shows that growth occurs to roughly the same size (in both populations) in the brown dwarf and T Tauri disks (at the same radial location with respect to the truncation radius).  This occurs because the estimates of the collisional timescales in units of the orbital timescales, $\tcoll/\torb$, are reasonably similar for the sizes that these aggregates grow to (Figure~\ref{fig:sigma_tcoll_BD_TT}, lower panel).  The coagulation and fragmentation kernels (given by $K_{ij}$ and $F_{ij}$, respectively) describe the rate at which particles of size $i$ and $j$ will collide and coagulate or fragment, respectively.  The collisional timescale is related to the kernels by:

\begin{equation}
\frac{\tcoll}{\torb} \propto \frac{\Omega}{(K_{ij}+F_{ij})N} \propto \frac{\Omega H}{(K_{ij}+F_{ij}) \Sigma},
\label{eq:tcoll}
\end{equation}
where $\Omega =\sqrt{G M_{\star}/R^3}$ is the angular frequency, $N$ is the number density of particles and $H =\cs/\Omega$ is the disk scaleheight.  Note that since $N$ changes with time, the latter proportionality uses the disk surface mass density and scaleheight as a proxy for the number density.  

\begin{figure}
\centering
  \includegraphics[width=1.0\columnwidth]{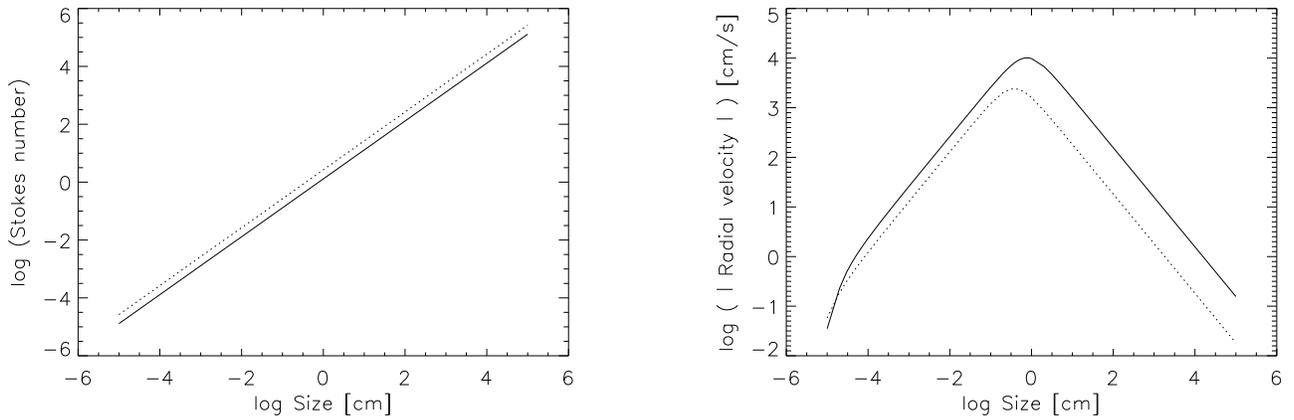}
\caption{The Stokes numbers for the brown dwarf disk at 10au (solid line) and the T Tauri disk at 60au (dotted line) are similar (within a factor of $\approx 2$), because the surface mass densities of the two disks are similar.}
 \label{fig:Stokes_BD_TT}
\end{figure}

Analytically, one can show that the collisional timescales are similar at the same location with respect to the truncation radius in both disks.  Using the aforementioned definitions for the scaleheight and the angular frequency, the observational relation $\Mdisc \propto \Rt^{1.6}$ \citep{Andrews_Md_Rt} and equations~\ref{eq:Sigma}, \ref{eq:cs} and~\ref{eq:tcoll}, we obtain

\begin{equation}
\frac{\tcoll}{\torb} \propto \left( \frac{R}{\Rt} \right )^{0.75} \frac{\Rt^{0.15} e^{R/\Rt}}{(K_{ij}+F_{ij})}.
\end{equation}
Hence, for any given disk location with respect to the outer radius, the collisional timescale has only a very weak dependence on the disk truncation radius and depends mainly on the collision kernels.  The collision kernels are goverened by the deterministic and stochastic velocities which are similar in both disks: for most of the relevant parameter space - i.e. particle sizes up to $\approx 10^2$ cm - the difference is less than an order of magnitude.  The stochastic velocities (equations 13, 19 and 29 of \citealp{Garaud_vel_pdf}) depend on the local sound speed and Stokes number (or equivalently, the surface mass density).  Using equations~\ref{eq:Sigma} and~\ref{eq:cs}, the ratios of the sound speeds and surface mass densities at equivalent locations in the brown dwarf and T Tauri disks are $c_{\rm s,BD}/ c_{\rm s,TT} \approx 1.6$ and $\Sigma_{\rm BD}/\Sigma_{\rm TT} \approx 2.3$, respectively.  The deterministic velocities (equations 14, 16 and 18 of \citealp{Garaud_vel_pdf}) depend on the Stokes number, the sound speed, the radial gas velocity, $u_{\rm g}$ (equation 41 of \citealp{Garaud_vel_pdf}), and $\eta v_{\rm k}$, where $\eta$ describes the deviation of the azimuthal gas velocity from the local Keplerian velocity, $v_{\rm k} = R \Omega$.  Using Section 4.1 of \citealp{Garaud_vel_pdf}, one can show that $\eta v_{\rm k} \propto \cs^2/(R \Omega)$ and that the ratio of this term at equivalent locations in the two disks is $(\eta v_{\rm k})_{\rm BD} / (\eta v_{\rm k})_{\rm TT} \approx 2.7$.  In addition, the ratio of the radial gas velocity at equivalent locations is $u_{\rm g, BD} / u_{\rm g, TT} \approx 4.2$.  Furthermore, Figure~\ref{fig:Stokes_BD_TT} shows the Stokes number (which affects the collisional dynamics since it affects all the velocities except that due to Brownian motion) against particle size for the brown dwarf disk at 10au and the T Tauri disk at 60au, and shows that these are within a factor of $\approx 2$ of each other.  Therefore, the factors that influence the deterministic and stochastic velocities are of order unity, thus indicating why the velocities are similar in both disks.  Consequently, the collisional timescales (in units of the orbital timescales) are similar, resulting in growth to similar maximum sizes in both disks.  The above results indicate that similar-sized particle distributions can exist in both T Tauri and brown dwarf disks if the latter are indeed smaller analogues of the former (which is the way these disks are set up).

Note the slight difference between the collisional timescales in the two discs, particularly for collisions between very small or very large particles (Figure~\ref{fig:sigma_tcoll_BD_TT}, lower panel).  This discrepancy arises because the disk properties and hence collision velocities, though very similar, are not \emph{exactly} the same.  The coagulation and fragmentation kernels are proportional to the mean sticking ($\bar{\epsilon}_{ij}^s$) and fragmentation ($\bar{\epsilon}_{ij}^f$) probabilities, respectively (equation 9 of \citealp{Garaud_vel_pdf}), the sum of which describe the probability that the collision results in an outcome other than bouncing.  The slightly different disk properties mean that the bouncing probability in the two disks will be slightly different for a collision between any two particle sizes because the collision velocities are not exactly the same.  This naturally leads to differences and hence an oscillation in the ratio of the collisional timescales.

We note that the simple formula for the collision timescale that is frequently used (i.e. $\tcoll \propto 1/(\Sigma \Omega)$) may not necessarily predict the \emph{location} at which growth will proceed to similar sizes in T Tauri and brown dwarf disks because it does not consider the local velocities that determine the complex collisional kernels (equations 32-34 of \citealp{Garaud_vel_pdf}).  Using the collision timescale of this simple form, the growth at 10au in the brown dwarf disk would be expected to be similar to that at 37au in our T Tauri disk as opposed to the 60au that we simulate.  However, the growth at 37au in the T Tauri disk does not match that in the brown dwarf disk at 10au: the position of the first peak of the surface mass density as a function of particle size - i.e. the characteristic size of the first particle population - is different.  In contrast, our method of considering grain growth at the same radial location with respect to the truncation radius yields results that are similar in both disks for both populations.

\section{Discussion}
\label{sec:disc}

\begin{figure}
  \centering
  \includegraphics[width=1.0\columnwidth]{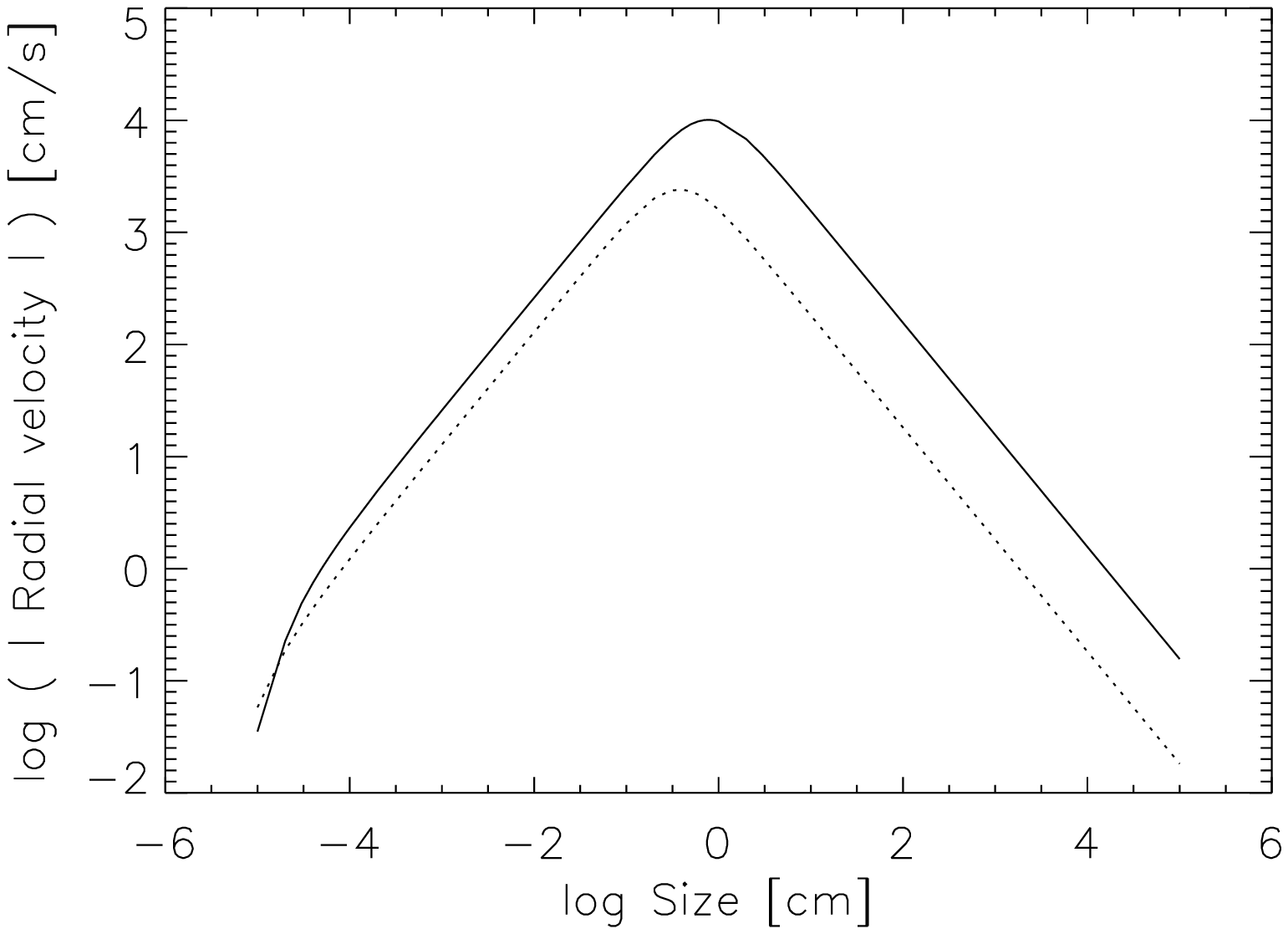}
  \caption{The radial drift velocities in the brown dwarf disk at 10au (solid line) and the T Tauri disk at 60au (dotted line) are similar for small particle sizes (within a factor of $\approx 2$) but deviate at larger sizes (by up to a factor of $\approx 10$).}
  \label{fig:uR_BD_TT}
\end{figure} 

Our results suggest that growth can proceed to similar sizes at equivalent locations with respect to the truncation radii in brown dwarf and T Tauri disks if the former are scaled-down versions of the latter.  This also suggests that the proportion of mass at any one particular size may well be similar.  If brown dwarf disks are indeed smaller versions of T Tauri disks our results suggest that the slope of the millimetre wavelength flux (when considering the total flux) may be expected to be the same in both disks.  This is supported by recent observations of brown dwarf disks which indicate that grain growth occurs to at least millimetre-sizes, similar to T Tauri disks \citep{Bouy_BDdiscs_mm,Ricci_mm_cm_BD_rhoOph,Ricci_mm_cm_2MASS,Mohanty_BD_VLMS_discs}.  Furthermore, if brown dwarf disks are scaled-down versions of T Tauri disks, spatially resolved brown dwarf disks may be expected to have spectral indices that change in similar ways to T Tauri disks - just on a smaller scale.

The radial velocities of dust aggregates in brown dwarf disks are expected to be larger than in T Tauri disks since (i) these are inversely proportional to the Keplerian velocity and hence increase with decreasing stellar mass, and (ii) for coupled particles (relevant for the start of dust growth) these increase with decreasing disk mass.  Therefore it is expected that grains will be lost into the central star.  Comparing the brown dwarf disk at 10au with the T Tauri disk at 60au, Figure~\ref{fig:uR_BD_TT} shows that the radial drift velocities are reasonably similar for small particles (within a factor of two).  Our model shows that growth is possible though we do not consider the radial flux of particles.  However, as yet, the importance of radial flux when considering the velocity PDFs of particles has not been explored so it is unclear what impact this would have on the loss of particles and hence the local grain size distribution - especially for larger sizes since this is where the radial velocities between the two disks deviate the most.  For an accurate view of whether particles will grow quickly beyond the radial drift barrier, such simulations should be performed.  In principle, if particles can grow sufficiently fast and overcome the radial drift barrier, we show that growth in brown dwarf disks may occur in similar ways to that in T Tauri disks if the former are scaled-down versions of the latter.  However, since the resolution has previously been shown to affect the growth timescales \citep{Garaud_vel_pdf} it is not possible to estimate if grains can grow past the radial drift barrier quickly enough.  Therefore our results show that \emph{in principle} growth in brown dwarf disks can occur to similar sizes as in T Tauri disks.

Furthermore, since growth occurs to similar sizes due to similar collisional timescales, if the outcome of one simulation is known, the estimates of the ratio of the collision timescales of the simulations may predict how growth will proceed in another simulation.

\begin{figure}
  \centering
  \includegraphics[width=1.0\columnwidth]{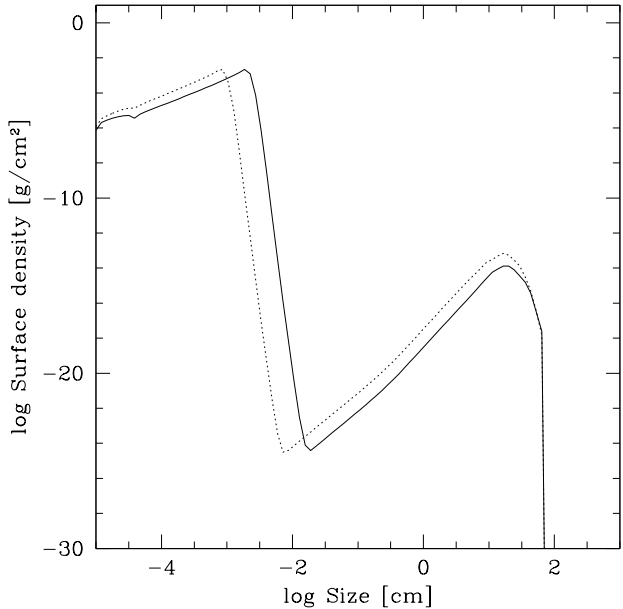}
  \caption{Surface mass density of aggregates against particle size for the T Tauri disk with sound speeds of $c_{\rm s,au} = 6.5 \times 10^4$cm/s (solid line) and $1.0 \times 10^5$cm/s (dotted line).  The temperature affects the absolute number density and the maximum size in the first population but not the maximum size.}
  \label{fig:TT_T_comp}
\end{figure}

To easily understand the effects of certain parameters on a many parameter problem, we use the same sound speed at 1au in both disks.  However, since we expect T Tauri disks to be hotter than brown dwarf disks at any one radius, we show the effects of increasing the sound speed at 1au to $c_{\rm s, au} = 1.0 \times 10^5$cm/s (i.e. as used by \citealp{Garaud_vel_pdf} and \citealp{Windmark_vel_pdf} when simulating growth with T Tauri disk parameters).  Figure~\ref{fig:TT_T_comp} shows that while growth is hindered in the first particle population, the maximum size is unaffected.

Finally, it has been suggested that growth may occur in brown dwarf disks (with particular disk conditions) if pressure inhomogeneities are present \citep{Pinilla_BD_discs}.  Our results show the potential for growth to occur even without the presence of these.

\section{Conclusions}
\label{sec:conc}

We perform coagulation and fragmentation simulations of dust growth in brown dwarf disks and show that growth occurs to large sizes.  We show that if brown dwarf disks are scaled-down versions of T Tauri disks (in terms of the stellar mass, disk mass and disk radius) growth can potentially occur up to similar sizes as in T Tauri disks at the same location relative to the disk truncation radius.

Recent observational results have shown that grain growth in brown dwarf disks can indeed occur up to millimetre-sizes, as observed in T Tauri disks.  In a first step, we show that our model has the potential to grow grains to large sizes in brown dwarf disks, in agreement with these recent observational findings.

\acknowledgments
We thank the referee and Pascale Garaud for useful comments, and the ISIMA2011 program at KIAA, Beijing.  FM is supported by ETH Zurich Postdoctoral Fellowship/Marie Curie Actions for People COFUND.  CO appreciates funding by DFG grant OL350/1-1.


\begin{thebibliography}{}

\bibitem[Andrews et al. 2010]{Andrews_Md_Rt} {{Andrews}, S.~M. and {Wilner}, D.~J. and {Hughes}, A.~M. and   {Qi}, C. and {Dullemond}, C.~P.}, 2010, \apj, 723, 1241

\bibitem[Apai et al. 2004]{Apai_BDdiscs_2micron} {{Apai}, D. and {Pascucci}, I. and {Sterzik}, M.~F. and {van der Bliek}, N. and   {Bouwman}, J. and {Dullemond}, C.~P. and {Henning}, T.} 2004, \aap, 426, L53

\bibitem[Apai et al. 2005]{Apai2005_BDdiscs_Sci} {Apai}, D. and {Pascucci}, I. and {Bouwman}, J. and {Natta}, A. and {Henning}, T. and {Dullemond}, C.~P.  2005 \Sci 310 834

\bibitem[Bouy et al. 2008]{Bouy_BDdiscs_mm}{{Bouy}, H. and {Hu{\'e}lamo}, N. and {Pinte}, C. and {Olofsson}, J. and   {Barrado Y Navascu{\'e}s}, D. and {Mart{\'{\i}}n}, E.~L. and   {Pantin}, E. and {Monin}, J.-L. and {Basri}, G. and {Augereau}, J.-C. and   {M{\'e}nard}, F. and {Duvert}, G. and {Duch{\^e}ne}, G. and   {Marchis}, F. and {Bayo}, A. and {Bottinelli}, S. and {Lefort}, B. and   {Guieu}, S.}, 2008, \aap, 486, 877

\bibitem[Galvagni et al. 2011]{Galvagni2011} {{Galvagni}, M. and {Garaud}, P. and {Meru}, F. and {Olczak}, C.}, 2011, Proceedings of the International Summer Institute for Modeling in Astrophysics, http://escholarship.org/uc/item/7n09f14x

\bibitem[Garaud et al. 2013]{Garaud_vel_pdf} Garaud, P., Meru, F., Galvagni, M. \& Olczak, C.  2013,  \apj, 764, 146

\bibitem[Klein et al. 2003]{Klein_BDdiscs_mm} {{Klein}, R. and {Apai}, D. and {Pascucci}, I. and {Henning}, T. and  {Waters}, L.~B.~F.~M.}, 2003, \apjl, 593, L57

\bibitem[Melzak 1957]{Melzak1957} {{Melzak}, Z. A.}, 1957, {{Trans. Amer. Math. Soc.}}, 85, 547

\bibitem[Mohanty et al. 2013]{Mohanty_BD_VLMS_discs} {{Mohanty}, S. and {Greaves}, J. and {Mortlock}, D. and {Pascucci}, I. and {Scholz}, A. and {Thompson}, M. and {Apai}, D. and {Lodato}, G. and {Looper}, D.}, 2013, ArXiv e-prints, 1305.6896

\bibitem[Okuzumi et al. 2011]{Okuzumi2011}{{Okuzumi}, S. and {Tanaka}, H. and {Takeuchi}, T. and {Sakagami}, M.-a.  }, 2011, \apj, 731, 95

\bibitem[Pinilla et al. 2013]{Pinilla_BD_discs} {{Pinilla}, P. and {Birnstiel}, T. and {Benisty}, M. and {Ricci}, L. and   {Natta}, A. and {Dullemond}, C.~P. and {Dominik}, C. and {Testi}, L.  }, 2013, ArXiv e-prints, 1304.6638

\bibitem[Ricci et al. 2012]{Ricci_mm_cm_BD_rhoOph} {{Ricci}, L. and {Testi}, L. and {Natta}, A. and {Scholz}, A. and   {de Gregorio-Monsalvo}, I.}, 2012, \apjl, 761 L20

\bibitem[Ricci et al. 2013]{Ricci_mm_cm_2MASS} {{Ricci}, L. and {Isella}, A. and {Carpenter}, J.~M. and {Testi}, L.  }, 2013, \apjl, 764, L27

\bibitem[Scholz et al 2006]{Scholz_BDdiscs_scaleddown_TT} {{Scholz}, A. and {Jayawardhana}, R. and {Wood}, K.}, 2006, \apj, 645, 1498

\bibitem[Shakura \& Sunyaev 1973]{SS_viscosity} {{Shakura}, N.~I. and {Sunyaev}, R.~A.} 1973, \aap, 24, 337

\bibitem[Smoluchowski 1916]{Smoluchowski1916} {{Smoluchowski}, M.~V.} 1916, {Zeitschrift fur Physik}, 17, 557

\bibitem[Teiser \& Wurm 2009]{Teiser_Wurm_decimeter_growth} {{Teiser}, J. and {Wurm}, G.} 2009, \aap, 505, 351

\bibitem[Teiser \& Wurm 2009]{Teiser_Wurm_highVcoll}{{Teiser}, J. and {Wurm}, G.}, 2009, \mnras, 393, 1584

\bibitem[Walker et al. 2004]{Walker_BDdisc_structure} {{Walker}, C. and {Wood}, K. and {Lada}, C.~J. and {Robitaille}, T. and   {Bjorkman}, J.~E. and {Whitney}, B.}, 2004, \mnras, 351, 607

\bibitem[Windmark et al. 2012a]{Windmark_lucky_ptcl} {{Windmark}, F. and {Birnstiel}, T. and {G{\"u}ttler}, C. and {Blum}, J. and {Dullemond}, C.~P. and {Henning}, T.}, 2012, \aap, 540, A73

\bibitem[Windmark et al 2012b]{Windmark_vel_pdf} {{Windmark}, F. and {Birnstiel}, T. and {Ormel}, C.~W. and {Dullemond}, C.~P.  }, 2012, \aap, 544, L16

\bibitem[Wurm et al. 2005]{Wurm_25m/s_impacts} {{Wurm}, G. and {Paraskov}, G. and {Krauss}, O.}, 2005, \icarus, 178, 253


\end{thebibliography}

\end{document}